 \documentclass[aps,prb,showpacs,groupedaddress]{revtex4}

\usepackage{graphicx}
\usepackage{bm}

\begin{document}

\preprint{}
\title{Formula for the Absorption Coefficient for Multi-Wall Nanotubes}
\author{Godfrey Gumbs}
\email{ggumbs@hunter.cuny.edu}
\author{  Antonios Balassis}
\email{abalassis@gmail.com}

\affiliation{Department of Physics and Astronomy,
Hunter College at the City University of New York, \\
695 Park Avenue New York, NY 10065}
\date{\today}

\begin{abstract}

We present a formalism for calculating the absorption coefficient of a pair of coaxial tubules.  A spatially
nonlocal, dynamical self-consistent field theory is obtained by calculating the electrostatic potential
produced by the charge density fluctuations as well as the external electric field.  There are peaks in the
absorption spectrum arising from plasma excitations corresponding either to plasmon or particle-hole modes.
In this paper, we numerically calculate the plasmon contribution to the absorption spectrum when an external
electric field is applied. The number of peaks depends on the radius of the inner as well as outer tubule.
The height of each peak is determined by the plasmon wavelength and energy.  For a chosen wave number, the
most energetic plasmon has the highest peak corresponding to the largest oscillator strength of the excited
modes. Some of the low-frequency plasmon modes have such weak coupling to an external electric field that
they are not seen on the same scale as the modes with larger energy of excitation. We plot the peak positions
of the plasmon excitations for a pair of coaxial tubules. The coupled modes on the two tubules are split by
the Coulomb interaction. The energies of the two highest plasmon branches increase with the radius of the
outer tubule. On the contrary, the lowest modes decrease in energy as this radius is increased. No effects
due to inter-tubule hopping are included in these calculations.
\end{abstract}


\maketitle

\section{Introduction}
\label{sec1}

Recently, there has been a considerable amount of effort
to understand the spectroscopic properties of
carbon nanotubes.\cite{new1a,new1,new2,new3,new4,new5,new6,new7}
In spectroscopic experiments, an external probe consisting
of a beam of, for example,  photons or electrons,
perturbs the structure. Consequently, the ensuing charge
density perturbation may be sustained by either single-particle
excitations or collective plasmon modes. The
density response may thus be measured through electron
energy loss, photoemission, or absorption experiments,
for example. So far, both theorists and experimentalists
have investigated the plasma excitations of nanotubes
using electron energy loss
spectroscopy (EELS).\cite{new8,new9,new10,new11,new12,new13,new14,new15,new16}
The theoretical formulation of EELS for nanotubes is now well
established and several predictions concerning plasmon
instabilities for nanotubes have been put forward.\cite{new8}
In this paper, we present a theory for the absorption coefficient
for a pair of coaxial tubules. The procedure can be generalized to multi-wall
nanotubes.
Several authors\cite{new1,new2,new3} have reported results for the optical
absorption spectra of single-wall carbon nanotubes. The role
played by the Coulomb interaction  enters through the
dielectric function. This was obtained for single-wall nanotubes using
the random-phase approximation (RPA) as well as time-dependent density
functional theory.\cite{new2,new3,new4} We separate the contributions
to the absorption coefficient due to plasmon and single-particle
excitations by carefully examining the dielectric function.

Our formalism relies on the use of the
nonlocal dynamical response function. The reason is that this
quantity determines the normal modes of oscillation which are identified
as resonances. It is also responsible for
the dielectric screening. We are able to separate
the contributions from single-particle and plasmon excitations.\cite{new17}
For simplicity, we use a continuum model to formulate
our theory for the absorption coefficient. However, the formula which we develop
may be adapted to nanotubes with bandstructures obtained using a basis
set of localized orbitals.\cite{new4}
We employ our results for the single-particle
eigenstates to determine the frequency-dependent nonlocal
dielectric function.
We neglect disorder by assuming that we are dealing with
high-quality nanotubes. For single-wall carbon nanotubes,
the mean-free paths typically exceed 1$\mu$m.
We use Maxwell's equations for the electrostatic potential
to calculate the dielectric response function of an electron
gas on a nanotube in the
RPA.\cite{new6}  Our analysis is specific to a double-wall
 nanotube but it can be extended in a straightforward way to a
multi-wall nanotubes.\cite{new7}


Zhang, et al. \cite{chao} calculated the dielectric
response function for single-wall carbon nanotubes. These authors
included the effect due to weak impurity scattering. Using their
dielectric response function and our formula derived below, we
would be able to identify the plasmon and single-particle excitations
for multi-wall nanotubes with both electron-electron and electron-impurity
scattering. Recently, the effects due to inter-tubule atomic hoppings and  inter-tubule electron-electron interactions
were included in calculations of the low-frequency electronic excitations of  double-wall armchair carbon
nanotubes.\cite{PLA} It was demonstrated in Ref.\ \cite{PLA} that the inter-tubule hopping significantly
modifies the low-frequency plasmon excitations. For example, there are additional  single-particle excitation
channels and tunneling plasmon modes. We do include these effects arising from tunneling. However, within the
framework of our calculations, these effects can be taken into account through the electron energy bandstructure.

The rest of this paper is organized as follows. In Sec.\ \ref{sec2},
we use a self-consistent field theory involving Poisson's equation
and linear response of the charge density fluctuations to
determine the induced potential. This potential is employed in the
absorption coefficient whose resonances are determined by the
plasma dispersion formula.  We must separate the contributions
 from the single-particle and plasmon excitations when calculating
the absorption coefficient. Gaps in the absorption coefficient
for single-particle excitations correspond to pockets
in the dispersion relation where the plasmons are not Landau damped.
Numerical results demonstrating these assertions are presented
in Sec.\ \ref{sec3}. Concluding remarks are given in Sec.\ \ref{sec4}.

\section{General Formulation of the Problem}
\label{sec2}

We now present a calculation for the absorption coefficient
when an incident beam of light of frequency $\omega$ interacts weakly
with a pair of coaxial tubules. In this case, linear response
theory is applicable for calculating the rate at which energy
is absorbed by the quantum system.  The formalism we present below
may be generalized to multi-wall cylindrical nanotubes consisting of an
arbitrary number of coaxial tubules and to an
array of parallel nanotubes, embedded in a material with dielectric
constant $\epsilon_b$. We use
Fermi's golden rule for the rate at which energy is
absorbed and then apply the fluctuation-dissipation
theorem to the density-density correlation function.\cite{callaway}
After some algebra, we find that the absorption coefficient
is given in terms of the real ($\Re e$) and imaginary ($\Im m$) parts
of the Lorentz ratio $\alpha_{\rm Lorentz}(\omega)$. This is defined
as the ratio of the Fourier coefficient of the absorbed energy at
frequency $\omega$ of a probing field to the square of its amplitude
and we have\cite{haug}

\begin{equation}
\beta_{\mbox{abs}}(\omega)=\frac{\omega}{n(\omega)\varepsilon_bc}
\left[1+\rho_{\mbox{ph}}(\omega)
\right]\ \Im m\  \alpha_{\rm Lorentz}(\omega)\ ,
\label{A1}
\end{equation}
where $\rho_{\mbox{ph}}(\omega)=\left[ \exp(\hbar\omega/k_BT)-1 \right]^{-1}$
is the photon distribution function, the refractive index is

\begin{equation}
n(\omega)= \left\{\frac{1}{2}\left[\epsilon_b+\Re e \ \alpha_{\rm Lorentz}(\omega)+
\sqrt{\left(\epsilon_b+\Re e \ \alpha_{\rm Lorentz}(\omega)  \right)^2+
\left( \Im m \ \alpha_{\rm Lorentz}(\omega) \right)^2}
\right]
\right\}^{1/2}
\ ,
\label{A2}
\end{equation}
and

\begin{equation}
\alpha_{\rm Lorentz}(\omega)=e\int d{\bf r}\
\delta n({\bf r},\omega)
\frac{\phi_{\mbox{ext}}\left({\bf r}\right)}{\left|\nabla
\phi_{\mbox{ext}}\left({\bf r}\right)  \right|^2}
\ .
\label{A3}
\end{equation}
 Here, $\phi_{\mbox{ext}}\left({\bf r}\right)$ is an arbitrary external
potential.  However, for   a uniform electric field, we have
$\phi_{\mbox{ext}}\left({\bf r}\right)=-{\bf r}\cdot
{\bf E}^{\mbox{ext}}$ for an electric field in the direction
with unit vector
$\hat{{\bf e}}_0={\bf E}^{\mbox{ext}}/
\left| {\bf E}^{\mbox{ext}}\right|$.
Also, the electron density fluctuation $\delta n({\bf r},\omega)$ on
the two tubules of inner radius $R_1$ and outer radius $R_2$ arises
from {\em both\/} the external electric field as well as the
electric field from the induced potential, i.e., in linear response
theory within the RPA, we have

\begin{equation}
\delta n({\bf r},\omega)\ = -e\int d{\bf r}^\prime\
\sum_{j=1}^2\chi_j^0\left({\bf r},
{\bf r}^\prime;\omega\right)\Phi_{\mbox{tot}}\left({\bf r}^\prime,
\omega\right)\ ,
\label{A4}
\end{equation}
where $\Phi_{\mbox{tot}}\left({\bf r},
\omega\right)=\phi_{\mbox{ext}}\left({\bf r}\right)+
\phi_{\mbox{ind}}\left({\bf r},\omega\right)$ is the sum of the
external potential and the induced potential
$\phi_{\mbox{ind}}\left({\bf r},\omega\right).$ The response function
for the j$^{\rm th}$ tubule is given by

\begin{eqnarray}
 \chi_j^0\left({\bf r},
{\bf r}^\prime;\omega\right)&=&2\sum_{\nu,\nu^\prime}
\frac{f_0(\epsilon_{j,\nu^\prime})-f_0(\epsilon_{j,\nu})}
{\hbar\omega+\epsilon_{j,\nu^\prime}-\epsilon_{j,\nu}}
\Psi_{j,\nu}^\ast({\bf r}^\prime)
\Psi_{j,\nu^\prime}({\bf r}^\prime)
\Psi_{j,\nu^\prime}^\ast({\bf r})
\Psi_{j,\nu}({\bf r})
\nonumber\\
&\equiv&\sum_{\nu,\nu^\prime} \Pi^0_{j,\nu\nu^\prime}(\omega) \Psi_{j,\nu}^\ast({\bf r}^\prime)
\Psi_{j,\nu^\prime}({\bf r}^\prime) \Psi_{j,\nu^\prime}^\ast({\bf r}) \Psi_{j,\nu}({\bf r}) \ . \label{A5}
\end{eqnarray}
In this notation, $f_0(\epsilon)$ is the Fermi-Dirac distribution function,
$\epsilon_{j,\nu}$ is an energy eigenvalue on the j$^{th}$
tubule and $\Psi_{j,\nu}({\bf r})$ the corresponding eigenfunction.
If an electron with effective mass $m^\ast$
is confined on the surface of a tubule
of radius $R_j$, whose axis is taken to be on the $z$ axis,
the eigenfunctions and eigenenergies are

\begin{mathletters}
\label{gae1}
\begin{equation}
\Psi_{j,\nu}(r,\varphi,z)=\frac{1}{\sqrt{L_z}}e^{ik_zz}\Phi_j(r) \frac{1}{\sqrt{2\pi R_j}}e^{il\varphi}\ ,
\Phi_j^2(r)=\delta(r-R_j)\ , \label{gae1a}
\end{equation}

\begin{equation}
\epsilon_{j,\nu}=\frac{\hbar^2k_z^2}{2m^{\ast}}
+\frac{\hbar^2l^2}
{2m^{\ast}R_j^2}
\ ,
\label{gae1b}
\end{equation}
\end{mathletters}
with $\displaystyle{\nu=\{k_z,l\},\ k_z=\frac{2\pi n}{L_z}}$,
$n,l=0,\pm 1,\pm 2,\cdots$ and $L_z$ is a normalization length.
Plasmon excitations can be obtained by
calculating the density matrix from its equation of motion
$\displaystyle{i\hbar\frac{\partial\hat{\rho}}
{\partial t}=[H,\hat{\rho}]}$. For small perturbations,
$\hat{\rho}=\hat{\rho}_0+\hat{\rho}_1$ and $H=H_0-e\phi$, where
$\hat{\rho}_0$ is the equilibrium density matrix and
$\hat{\rho}_1$ its
perturbation, $H_0$ is the unperturbed Hamiltonian and $\phi(r,\varphi,z;t)$
is the fluctuation in the electric potential corresponding to
$\hat{\rho}_1$.

After we substitute Eq.\ (\ref{A5}) into Eq.\ (\ref{A4}), we obtain the
following result for Eq.\ (\ref{A3}), i.e.,

\begin{equation}
\alpha_{\rm Lorentz}(\omega)=-e^2 \sum_{j,\nu,\nu^\prime}
\Pi^0_{j,\nu\nu^\prime}(\omega)\left\{Q_{j,\nu\nu^\prime}+U_{j,\nu\nu^\prime}(\omega)
 \right\} \ M_{j,\nu^\prime\nu} \ ,
\label{A6}
\end{equation}
where

\begin{equation}
M_{j,\nu\nu^\prime}= \int d{\bf r}\ \frac{\phi_{\mbox{ext}}\left({\bf r}\right)}
{\left|\nabla
\phi_{\mbox{ext}}\left({\bf r}\right)  \right|^2}\
\Psi_{j,\nu}^\ast({\bf r})\Psi_{j,\nu^\prime}({\bf r})\ ,
\label{A7}
\end{equation}

\begin{equation}
Q_{j,\nu\nu^\prime}=        \int d{\bf r}\ \phi_{\mbox{ext}}\left({\bf r}\right)\
\Psi_{j,\nu}^\ast({\bf r})\Psi_{j,\nu^\prime}({\bf r})\
\label{A7.2}
\end{equation}
are transition matrix elements and we have also introduced

\begin{equation}
U_{j,\nu\nu^\prime}(\omega)=\int d{\bf r}\
\phi_{\mbox{ind}}({\bf r},\omega)
\
\Psi_{j,\nu}^\ast({\bf r})\Psi_{j,\nu^\prime}({\bf r})\ .
\label{A8}
\end{equation}

In the presence of an external potential $\phi_{\mbox{ext}}$,
the induced potential is a solution of Poisson's equation

\begin{equation}
{\bf\nabla}^2\phi_{\mbox{ind}}({\bf r},\omega) =\frac{4\pi
e}{\epsilon_s} \left(\delta n_{\mbox{ind}}({\bf r},\omega) -e\int
d{\bf r}^\prime\sum_{j=1}^2 \ \chi_j^0({\bf r},{\bf
r}^\prime;\omega)\ \phi_{\mbox{ext}}({\bf r}^\prime)  \right)\ .
\end{equation}
The electrostatic potential $\phi_{\mbox{ind}}(r,\varphi,z;\omega)$
can be Fourier transformed with respect to the variables $\varphi$ and $z$ to
give

\begin{equation}
\phi_{\mbox{ind}}(r,\varphi,z;\omega)=\sum_{q_z,L}
\phi_{\rm L}(r,q_z,\omega)e^{iq_zz}e^{iL\varphi}
\ ,\ q_z=\frac{2n\pi}{L_z}\ ,
\ \ n,L=0,\pm 1,\pm 2,\cdots\ ,
\label{A9}
\end{equation}
so that Poisson's equation becomes

\begin{eqnarray}
& &\frac{1}{r}\frac{\partial}{\partial r}\left(r \frac{\partial
\phi_{\rm L}(r,q_z,\omega)}{\partial r} \right)-\frac{L}{r^2}
\phi_{\rm L}(r,q_z,\omega)-q^2
\phi_{\rm L}(r,q_z,\omega)
\nonumber\\
&=&\sum_{j=1}^2\left(A_{\rm L}^{(j)}(q_z,\omega)+B_{\rm L}^{(j)}(q_z \ , \omega)\right)\delta(r-R_j)\ .
\label{A10}
\end{eqnarray}
Here,

\begin{eqnarray}
A_{\rm L}^{(j)}(q_z,\omega)&=&\frac{-2e}{\varepsilon_s R_j}
\sum_{l=-\infty}^{\infty} \int_{-\infty}^{\infty}dk_z\ \langle
j,k_zl|\hat{\rho}_1|j,k_z-q_z,l-L\rangle
\nonumber\\
&=&\frac{2e^2}{\varepsilon_s R_j}\phi_{\rm L}(R_j,q_z,\omega)
\chi_{\rm j,L}(q_z,\omega) \label{A11}
\end{eqnarray}
and $\hat{\rho}_1$ denotes the perturbed density matrix to lowest
order in the external perturbation.
The polarization function of the electron gas on each
nanotube is defined by

\begin{eqnarray}
\chi_{\rm j,L}(q_z,\omega)&=&2\sum_{l,l^\prime=-\infty}^\infty
\int_{-\infty}^\infty \frac{dk_z}{2\pi}
\frac{f_0(\epsilon_{j,k_z,l})-
f_0(\epsilon_{j, k_z-q_z,l-L})}{\hbar\omega+
\epsilon_{j,k_z-q_z,l-L}
-\epsilon_{j,k_z,l}}
\nonumber\\
&=&\sum_{l=-\infty}^\infty \int_{-\infty}^\infty \frac{dk_z}{2\pi} \Pi^0_{j}(k_z,l;k_z-q_z,l-L) \ .
\label{A12}
\end{eqnarray}
Also, in Eq.\ (\ref{A10}), we have introduced the quantity
$B_{\rm L}^{(j)}(q_z,\omega)$ which depends on the external
electric field as follows

\begin{eqnarray}
B_{\rm L}^{(j)}(q_z,\omega) &=& \frac{-2e}{\varepsilon_s R_j}
\sum_{l=-\infty}^{\infty} \int_{-\infty}^{\infty}dk_z  \
\Pi^0_{j}(k_z,l;k_z-q_z,l-L)
\nonumber\\
&\times&
\int d{\bf r}\ \Psi_{j,k_z,l}^\ast ({\bf r})
\Psi_{j,k_z-q_z,l-L}\phi_{\mbox{ext}}({\bf r})\ .
\label{A14}
\end{eqnarray}

We now write the solution of Poisson's equation for
$\phi_L(r,q_z,\omega)$ in Eq.\ (\ref{A10}) as

\begin{equation}
\phi_L(r,q_z,\omega)=\left\{ \matrix{
\overline{C}_1I_L(q_zr),& r <R_1\cr
\overline{C}_3I_L(q_zr)+\overline{C}_4K_L(q_zr),& R_1\leq r\leq R_2\cr
 \overline{C}_2K_L(q_zr), & r>R_2\cr
}\right.\ ,
\label{gae29}
\end{equation}
where $I_L(x)$ and $K_L(x)$ are Bessel functions of imaginary argument and
$\overline{C}_1$, $\overline{C}_2$, $\overline{C}_3$ and $\overline{C}_4$
are functions of $q_z$ and $\omega$ but are independent of the
radial coordinate $r$.  These coefficients
are determined from the continuity of
$\phi_L(r,q_z;\omega)$ at $r=R_1,R_2$ and the step-like change of
$\displaystyle{\frac{\partial \phi_L}{\partial r}}$ at $r=R_1,R_2$, obtained
by integrating  Eq.\ (\ref{A10}) across the surfaces of the tubules.
These boundary conditions together yield

\begin{equation}
\left( \matrix{ I_L(q_zR_1)&0&-I_L(q_zR_1)&-K_L(q_zR_1)\cr 0&-K_L(q_zR_2)&I_L(q_zR_2)& K_L(q_zR_2)\cr
-I_L^{\prime}(q_zR_1)&0& I_L^{\prime}(q_zR_1)& K_L^{\prime}(q_zR_1)\cr 0&
K_L^{\prime}(q_zR_2)&-I_L^{\prime}(q_zR_2)& -K_L^{\prime}(q_zR_2)\cr }\right) \left(\matrix{
\overline{C}_1\cr \overline{C}_2\cr \overline{C}_3\cr \overline{C}_4\cr} \right)=\left(\matrix{0\cr 0\cr
\frac{A_L^{(1)}(q_z,\omega)+B_L^{(1)}(q_z,\omega)}{q_z}\cr
\frac{A_L^{(2)}(q_z,\omega)+B_L^{(2)}(q_z,\omega)}{q_z}\cr}\right) \ . \label{gae30}
\end{equation}
In this notation, the prime on $I_L(x)$ and $K_L(x)$ indicates
that a derivative with respect to the argument must be taken.
This set of simultaneous equations
 show that $\overline{C}_1,\ \overline{C}_2,\
\overline{C}_3$ and $\overline{C}_4$ are determined by the induced
potential through $A_L^{(j)}(q_z,\omega),\ B_L^{(j)}(q_z,\omega)$
on the right-hand side. On the other hand,
the induced potential is
determined by $\overline{C}_j$ in Eq.\ (\ref{gae29}). This self-consistent
field theory allows us to obtain the following analytic results for the
coefficients as

\begin{eqnarray}
& &\overline{C}_1(q_z,\omega)=-R_1K_L(q_zR_1)
\nonumber\\
&\times&\left[A_L^{(1)}(q_z,\omega)+B_L^{(1)}(q_z,\omega)
\right]-R_2K_L(q_zR_2)\left[A_L^{(2)}(q_z,\omega)+B_L^{(2)}(q_z,\omega)
\right]\ ,
\label{gae31a}
\end{eqnarray}

\begin{eqnarray}
& &\overline{C}_2(q_z,\omega)=-R_1I_L(q_zR_1)
\nonumber\\
&\times&\left[A_L^{(1)}(q_z,\omega)+B_L^{(1)}(q_z,\omega)
\right]-R_2I_L(q_zR_2)\left[A_L^{(2)}(q_z,\omega)+B_L^{(2)}(q_z,\omega)
\right]\ .
\label{gae31b}
\end{eqnarray}

\begin{equation}
\overline{C}_3(q_z,\omega)=-R_2K_L(q_zR_2)\left[A_L^{(2)}(q_z,\omega)+B_L^{(2)}(q_z,\omega)
\right]\ .
\label{gae31c}
\end{equation}

\begin{equation}
\overline{C}_4(q_z,\omega)=-R_1I_L(q_zR_1)\left[A_L^{(1)}(q_z,\omega)+B_L^{(1)}(q_z,\omega)
\right]\ .
\label{gae314}
\end{equation}
Making use of these results in Eq.\ (\ref{gae29}), we obtain
$\phi_L(R_1,q_z,\omega)=\overline{C}_1I_L(q_zR_1)$ and
$\phi_L(R_2,q_z,\omega)=\overline{C}_2K_L(q_zR_2)$ on
the surfaces of the tubules. Combining these results for $\phi_L(R_j,q_z,\omega)$
on the surface of the tubule with
Eq.\ (\ref{A11}) where $A_{\rm L}^{(j)}(q_z,\omega)$ is
given in terms of $\phi_{\rm L}(R_j,q_z,\omega)$, these self-consistent
field results yield

\begin{eqnarray}
& &A_{\rm L}^{(1)}(q_z,\omega)+B_{\rm L}^{(1)}(q_z,\omega)=
\nonumber\\
& &\frac{1}{{\cal D}_{\rm L}(R_1,R_2;q_z,\omega)}\left\{
\varepsilon_{\rm L}^{(2)}(q_z,\omega)B_{\rm L}^{(1)}(q_z,\omega)-
\frac{R_2}{R_1} V_{12}(q_z)\chi_{\rm 1,L}(q_z,\omega)
B_{\rm L}^{(2)}(q_z,\omega)
\right\}
\label{AB1}
\end{eqnarray}

\begin{eqnarray}
& &A_{\rm L}^{(2)}(q_z,\omega)+B_{\rm L}^{(2)}(q_z,\omega)=
\nonumber\\
& &\frac{1}{{\cal D}_{\rm L}(R_1,R_2;q_z,\omega)}\left\{ \varepsilon_{\rm L}^{(1)}(q_z,\omega)B_{\rm
L}^{(2)}(q_z,\omega)- \frac{R_1}{R_2} V_{12}(q_z)\chi_{\rm 2,L}(q_z,\omega) B_{\rm L}^{(1)}(q_z,\omega)
\right\} \ , \label{AB2}
\end{eqnarray}
where

\begin{eqnarray}
\varepsilon_{\rm L}^{(j)}(q_z,\omega)&=&
1+\frac{2e^2}{\varepsilon_s}I_L(qR_j)K_L(qR_j) \chi_{{\rm j,L}}
(q_z,\omega) \ ,
\nonumber\\
V_{12}(q_z)&=& \frac{2e^2}{\varepsilon_s}I_L(qR_1)K_L(qR_2)
\nonumber\\
{\cal D}_{\rm L}(R_1,R_2;q_z,\omega)&=& \varepsilon_{\rm L}^{(1)}(q_z,\omega) \varepsilon_{\rm
L}^{(2)}(q_z,\omega)-V_{12}^2(q_z) \chi_{{\rm 1,L}} (q_z,\omega)\chi_{{\rm 2,L}} (q_z,\omega)\ .
\label{gae24}
\end{eqnarray}
In the absence of an external perturbation, $B_L^{(j)}(q_z,\omega)=0$
and the nontrivial solutions for $A_L^{(j)}(q_z,\omega)$ in
Eqs.\ (\ref{AB1}) and (\ref{AB2}) correspond to
${\cal D}_{\rm L}(R_1,R_2;q_z,\omega)=0$

We may now express $M_{j,\nu\nu^\prime}$ in Eq.\ (\ref{A7}) and
$U_{j,\nu\nu^\prime}(\omega)$ in Eq.\ (\ref{A8}) in terms of Fourier  series
expansions with respect to the variables $\varphi$ and $z$. This was
done for the induced potential in Eq.\ (\ref{A9}). We then have

\begin{eqnarray}
M_{j,\nu\nu^\prime}&=&-\int d{\bf r}\ X\left({\bf r}\right)\
\Psi_{j,\nu}^\ast({\bf r})\Psi_{j,\nu^\prime}({\bf r})\
\nonumber\\
&=&-\sum_{q_z}\sum_{L=0,\pm 1,\pm 2,\cdots}\int d{\bf r} \
e^{iq_zz}e^{iL\varphi} \ \tilde{X}_L(k_z)\
\Psi_{j,\nu}^\ast({\bf r})\Psi_{j,\nu^\prime}({\bf r}) \ ,
\label{A7b}
\end{eqnarray}
where
$X\left({\bf r}\right)\equiv \phi_{\mbox{ext}}\left({\bf r}\right)/
\left|\nabla
\phi_{\mbox{ext}}\left({\bf r}\right)  \right|^2.$ We also use similar
Fourier expansions for $Q_{j,\nu\nu^\prime}$ and $U_{j,\nu\nu^\prime}$
in Eqs.\ (\ref{A7.2}) and (\ref{A8}), respectively.
When the eigenfunctions are substituted into Eq.\ (\ref{A7b}), then we find that
the only non-zero matrix elements occur when $l^\prime=l+L$
and $k_z^\prime=k_z+q_z$.

We can separate the contributions to the absorption
coefficient from plasmons and single-particle excitations.
The normal modes of oscillation for each tubule when the coupling
$V_{12}$ between them is neglected are obtained by solving
$\varepsilon_{\rm L}^{(j)}(q_z,\omega)=0$. Below,
we derive explicit results for the induced potential for a single tubule
of radius R.

For a single tubule of radius R, we write the Fourier component
of the induced potential as

\begin{equation}
\phi_{\rm L}(r,q_z,\omega)=
\left\{\matrix{C_1 I_L(q_zr), & r<R\cr
C_2 K_L(q_zr), & r>R\cr}
\right.\ .
\label{A15}
\end{equation}
After applying the boundary conditions on the surface of the
tubule to the continuity of the potential and the discontinuity
of its derivative, as seen in (\ref{A10}), we obtain

\begin{eqnarray}
C_1(q_z,\omega)&=& -\left[A_{\rm L}(q_z,\omega)+B_{\rm L}(q_z,\omega)  \right]RK_L(q_zR)
\nonumber\\
C_2(q_z,\omega)&=& -\left[A_{\rm L}(q_z,\omega)+B_{\rm L}(q_z,\omega)  \right]RI_L(q_zR)\ ,
\label{A16}
\end{eqnarray}
where $A_{\rm L}$ and $B_{\rm L}$ are as defined above for a tubule
of radius R. We now use the results in Eq.\ (\ref{A16}) to obtain the
electrostatic potential $\phi_{\rm L}(R,q_z,\omega)$ on the surface
of the tubule. When the derived result is substituted into
Eq.\ (\ref{A11}), we obtain

\begin{equation}
A_{\rm L}(q_z,\omega)+B_{\rm L}(q_z,\omega)=\frac{B_{\rm L}(q_z,\omega) }
{\varepsilon_{\rm L}(q_z,\omega)}\ ,
\label{A17}
\end{equation}
which can now be employed to calculate $\phi_{\rm L}(r,q_z,\omega)$
in all regions by making use of Eqs.\ (\ref{A15}) and (\ref{A16}).
Here, $\varepsilon_{\rm L}(q_z,\omega)$ is given by
$\varepsilon_{\rm L}^{(j)}(q_z,\omega)Eq.\ (\ref{gae24})$
for a tubule of radius $R_j=R$.

In the absence of an external electric field, we must set
$\phi_{\mbox{ext}}({\bf r})$ in Eq.\ (\ref{A14}) equal to zero.
This results in $B_{\rm L}=0$ so that Eq.\ (\ref{A17}) reduces to
$\epsilon_{\rm L}(q_z,\omega)A_{\rm L}(q_z,\omega)=0$, where
$\epsilon_{\rm L}(q_z,\omega) = 1+
(e^2/\epsilon_s)I_L(q_zR)K_L(q_zR)\chi_{\rm L}(q_z,\omega)$ is the
dielectric function for the nanotube. A nontrivial solution
for $A_{\rm L}(q_z,\omega)$ leads to the plasma
 dispersion equation $\epsilon_{\rm L}(q_z,\omega)=0$ for the
{\em self-sustaining\/} plasma oscillations. However, in the presence
of an external electric field, the solution of Eq.\ (\ref{A17}) for
$A_{\rm L}(q_z,\omega)$ in terms of $B_{\rm L}(q_z,\omega)$ leads to the
coefficients $C_1$ and $C_2$ in Eq.\ (\ref{A16}) as
$C_1=-RK_L(q_zR)B_{\rm L}(q_z,\omega)/\epsilon_{\rm L}(q_z,\omega)$ and
 $C_2=-RI_L(q_zR)B_{\rm L}(q_z,\omega)/\epsilon_{\rm L}(q_z,\omega)$. When these
results are substituted into Eq.\ (\ref{A15}), we obtain a
closed-form analytic result for $\phi_{\rm L}(r,q_z,\omega)$ and
consequently $\phi_{\mbox{ind}}(r,\varphi,z;\omega)$ in Eq.\ (\ref{A9})
by making use of the Fourier series expansion. We may now use these
results to obtain $U_{\alpha\alpha^\prime}(\omega)$  in Eq.\
(\ref{A8}) which could be employed in the Lorentz ratio
in Eq.\ (\ref{A6}) for calculating the absorption coefficient
$\beta_{\mbox{abs}}(\omega)$ in Eq.\ (\ref{A1}).

We now turn to analyzing the contributions to $\alpha_{\rm Lorentz}(\omega)$
in Eq.\ (\ref{A6}).  This means that we must examine when the factor
$\Im m\ \chi^0_{j,L}(\omega)\phi_{L}(R_j,q_z,\omega)\neq 0$. Thus, we must determine
$\Im m \ {\cal D}_{\rm L}(R_1,R_2;q_z,\omega)^{-1}
=-\frac{{\cal D}_{\rm I}}{{\cal D}_{\rm I}^2+{\cal D}_{\rm R}^2}$
where ${\cal D}_{\rm R}$ and ${\cal D}_{\rm I}$ are the real and imaginary parts
of ${\cal D}_{\rm L}$.
Therefore, the only non-zero
contributions to $\beta_{\mbox{abs}}(\omega)$ arise
when either ${\cal D}_{\rm I}$ is finite, i.e., from the single-particle
excitations, or when {\em both} ${\cal D}_{\rm R}$ and
${\cal D}_{\rm I}$ vanish simultaneously. This latter contribution comes
from the plasmon excitations which are not Landau damped by the
particle-hole modes.

\section{Numerical Results}
\label{sec3}

In Fig.\ \ref{fig2}, we plot the plasmon contribution to the
absorption coefficient when $L=0$ and $q_z=0.15\ k_F$ as a function
of the incident photon energy for inner radius $R_1=10.0$
\nolinebreak {\AA} and several values of the radius $R_2$ of the
outer tubule for a pair of coaxial tubules. In Fig.\ \ref{fig3}, the
peak positions of the two highest branches of plasmon excitation energies for a pair of
coaxial tubules are plotted as a function of $R_2/R_1$ for fixed
$R_1=10.0$ {\AA}. The parameters used to do our numerical
calculations were chosen as follows. The electron effective mass
$m^\ast=0.25\ m_e$, where $m_e$ is the free electron mass, the Fermi
energy $E_F=0.6$ eV and the background dielectric constant
$\epsilon_b=2.4$.\cite{new6}    Our calculations  show that the number
of peaks depends on  the chosen wave vector $q$
as well as the ratio $R_2/R_1$. However,   for the range $0\leq q/k_F\leq 0.2$, the two highest
branches are excited.    There are some small amplitude oscillations on the curve in Fig.\ \ref{fig3}
representing the highest plasmon excitation energy.  This behavior is caused by the variation in
the number of occupied energy levels on the outer cylinder as $R_2$ is changed for a fixed
number of electrons, i.e., $E_F$. The oscillations of the second highest branch are negligible
because the pocket in the single-particle excitation spectrum  where that branch exists is very narrow.
  The intensity (oscillator
strength) of each peak is determined by  the plasmon excitation energy.  Consistently,
the highest peak has the largest oscillator strength.\cite{new6}
Some of the less energetic plasmon modes have such weak coupling to
an external electric field that they are not seen on the scale we
used in Fig.\ \ref{fig2}.

In Fig. \ref{fig4}, we plot the particle-hole mode contribution to the absorption
for a pair of coaxial tubules. The curve in Fig.\ \ref{fig4} is discontinuous Since the imaginary part of
${\cal D}$ is zero in the pockets where the plasmon frequencies exist. The gaps on the $\omega$-axis
correspond to the pockets in the excitation spectrum where the collective plasmon modes exist. The dispersion
of plasmon and single-particle excitations is given in Ref.\ |cite{{new6}. , this accounts for the gaps on
the frequency axis of Fig.\ \ref{fig4}. We chose $q_z=0.15\ k_F$ and all other parameters for the electron
effective mass, Fermi energy and background dielectric constant are the same as Fig.\ \ref{fig2}. Unlike the
plasmon contribution in Fig.\ \ref{fig2}, the particle-hole modes contribute over a wider range of frequency.
The plasmon excitations occur within pockets between the particle-hole modes where they are Landau
damped.\cite{new6} In Fig. \ \ref{fig5}, we plot these particle-hole mode frequencies versus $q_z/k_F$ for
the pair of coaxial tubules used in Fig.\ \ref{fig4}. Numerical results for plasma excitations between
subbands for which $L=1,2,\cdots$ may also be obtained from our formula for the absorption coefficient.

\section{Concluding Remarks and Summary}
\label{sec4}

In this paper, we gave a comprehensive formalism for calculating
the absorption coefficient for a double-wall nanotube. The
calculation was carried out using a self-consistent field approach for
the induced potential and charge density fluctuations on the tubules.  Our
result is given in terms of the electron-electron interaction on
each tubule and between the two tubules. We did not include any effects due to inter-tubule atomic hopping.\cite{PLA}  The effective dielectric
function ${\cal D}_{\rm L}(R_1,R_2;q_z,\omega)$ for the pair of
tubules is expressed in terms of the dielectric functions
$\varepsilon_{\rm L}^{(1)}(q_z,\omega)$ and $\varepsilon_{\rm
L}^{(2)}(q_z,\omega)$ for each tubule, as shown in Eq.\
(\ref{gae24}). The electron gas model for each tubule was used to
simplify the calculations. However, a more realistic model whose
energy bands are obtained using a tight-binding approximation for
example, would be incorporated through the susceptibility
$\chi_{{\rm 1,L}} (q_z,\omega)$ and $\chi_{{\rm 2,L}} (q_z,\omega)$
for each tubule. We showed that the loss function Im $\left[1/{\cal
D}_{\rm L}(R_1,R_2;q_z,\omega)\right]$ can be separated into
contributions due to plasmon excitations and particle-hole modes.
Figures\ \ref{fig4} and \ref{fig5} show that there are pockets
within the particle-hole continuum where there is no Landau damping
of the collective plasmon excitations. These regions could only be
determined by separating the contributions to the loss function from
the plasmon and single-particle excitations, as we described above.
This separation would allow direct comparison between theory and
experimental results of the absorption spectrum for plasma
excitations on nanotubes.

\acknowledgments This work was supported by contract  FA 9453-07-C-0207 of AFRL.

\newpage

 \begin{figure}
   \begin{center}
   \begin{tabular}{c}
\includegraphics[width=10cm]{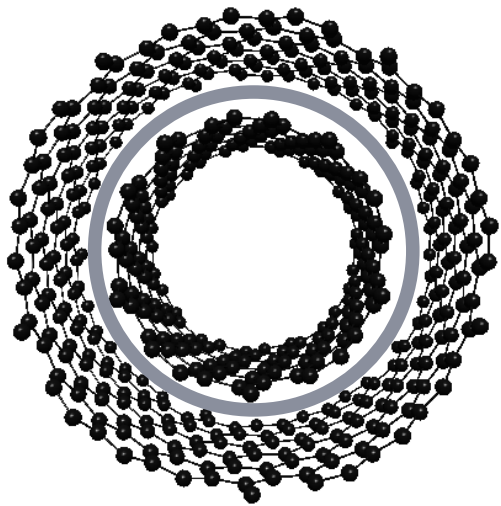}
   \end{tabular}
   \end{center}
   \caption[example]
   { \label{fig1}
Schematic of a pair of coaxial tubules. }
\end{figure}

   \begin{figure}
   \begin{center}
   \begin{tabular}{c}
 \includegraphics[width=10cm]{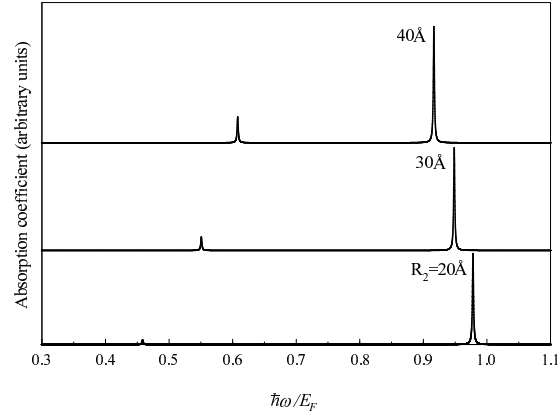}
   \end{tabular}
   \end{center}
   \caption[example]
   { \label{fig2}
The $L=0$ plasmon contribution to the absorption coefficient
(arbitrary units) versus photon energy for $q_z=0.15\ k_F$, where
$k_F=\sqrt{2m^\ast E_F/\hbar^2}$ with the radius of the inner tubule
$R_1=10.0$ {\AA} and several values of $R_2$. For clarity, the
curves are shifted. From the bottom, the curves correspond to $R_2
=20$ {\AA},\ 30 {\AA} and  40 {\AA}. The parameters used in the
calculation are given in the text for $m^\ast,\ E_F$ and
$\epsilon_b$.}
   \end{figure}

  \begin{figure}
   \begin{center}
   \begin{tabular}{c}
  \includegraphics[width=10cm]{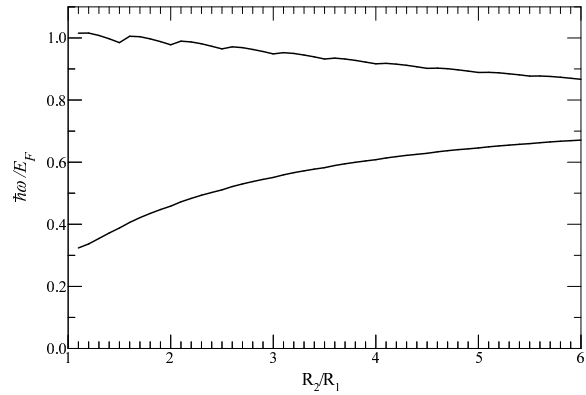}
   \end{tabular}
   \end{center}
   \caption[example]
   { \label{fig3}
Peak positions for the $L=0$ and $q_z=0.15\ k_F$ plasmon excitations
for the absorption coefficient for a pair of coaxial tubules as a
function of $R_2/R_1$, for inner radius $R_1=10.0$ {\AA} and outer
radius $R_2$. The parameters used in the calculation are given in
the text for $m^\ast,\ E_F$ and $\epsilon_b$ and are the same as
Fig.\ \ref{fig2}. }
  \end{figure}

    \begin{figure}
   \begin{center}
   \begin{tabular}{c}
 \includegraphics[width=10cm]{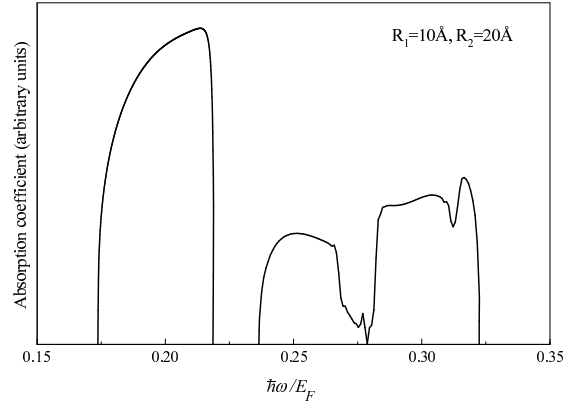}
   \end{tabular}
   \end{center}
   \caption[example]
   { \label{fig4}
The $L=0$ particle-hole mode contribution to the absorption
coefficient (arbitrary units) versus photon energy for $q_z=0.15\
k_F$, where $k_F=\sqrt{2m^\ast E_F/\hbar^2}$. The radius of the
inner tubule $R_1=10.0$ {\AA} and $R_2=20.0$ {\AA} is the radius of
the outer tubule . Only the frequency range where there are
particle-hole modes contributes. The gaps on the frequency axis
correspond to pockets where plasmon excitations are not Landau
damped. The parameters used in the calculation are given in the text
for $m^\ast,\ E_F$ and $\epsilon_b$.}
   \end{figure}

    \begin{figure}
   \begin{center}
   \begin{tabular}{c}
   \includegraphics[width=10cm, angle=-90]{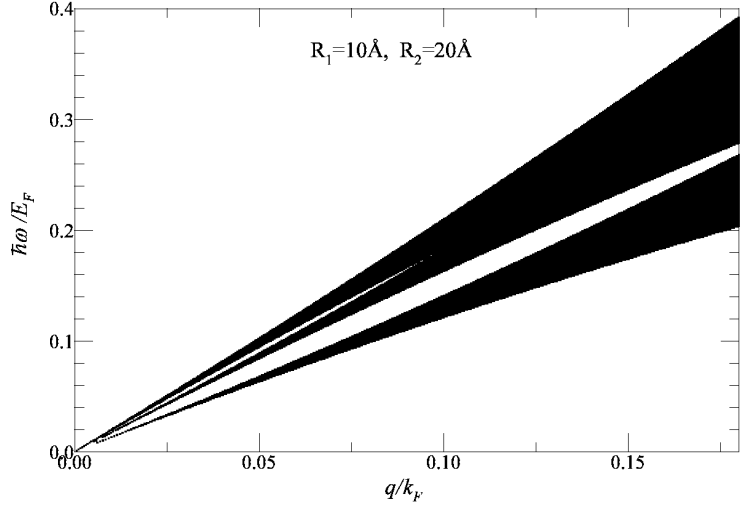}
   \end{tabular}
   \end{center}
   \caption[example]
   { \label{fig5}
The dispersion relation for the $L=0$ continuum of particle-hole modes (shaded regions)
for the pair of coaxial tubules used in Fig.\ \ref{fig4}. }
   \end{figure}

\end{document}